# Scalable Predictive Time-Series Analysis of COVID-19: Cases and Fatalities

Shradha Shinde [a], Jay Joshi [a], Sowmya Mareedu [a], Yeon Pyo Kim [b], Jongwook Woo [a]

[a]Department of Information System, California State University Los Angeles

[b]SoftZen Co. Ltd.

## Abstract

*COVID 19 is an acute disease that started spreading throughout the world, beginning in December 2019. It has spread worldwide and has affected more than 7 million people, and 200 thousand people have died due to this infection as of Oct 2020. In this paper, we have forecasted the number of deaths and the confirmed cases in Los Angeles and New York of the United States using the traditional and Big Data platforms based on the Times Series: ARIMA and ETS. We also implemented a more sophisticated time-series forecast model using Facebook Prophet API. Furthermore, we developed the classification models: Logistic Regression and Random Forest regression to show that the Weather does not affect the number of the confirmed cases. The models are built and run in legacy systems (Azure ML Studio) and Big Data systems (Oracle Cloud and Databricks). Besides, we present the accuracy of the models.*

## Introduction

COVID-19 is an infectious disease caused by a newly discovered coronavirus. Most people infected with the COVID-19 virus have experienced mild to moderate respiratory illness and sometimes recovered without requiring special treatment. However, senior people and those with underlying medical problems like cardiovascular disease, diabetes, chronic respiratory disease, and cancer are more likely to develop serious illnesses. The right way to prevent and slow down transmission is to be well informed about the COVID-19 virus, the disease it causes, and how it spreads. People can protect themselves and others from infection by wearing masks, washing hands, or using an alcohol-based rub frequently and not touching their faces.

It was first diagnosed in Wuhan, China, in late 2019. Since then, it has spread worldwide and has affected more than 7 million people, and 200 thousand people have died due to this infection as of October 2020. Scientists have since then dedicated their time to researching this disease, and here we attempt to take a more in-depth look into this problem using traditional and Big Data machine learning models.

We can define Big Data as non-expensive frameworks, mostly in distributed parallel computing systems, which can store a large-scale data and process it in parallel. A large-scale data means data of giga-bytes or more, which cannot be processed well or too expensive using traditional computing systems. It is also linearly scalable (Woo & Xu, 2011, 2013).

In the Big Data community, the legacy MapReduce's scheduling overhead and lack of support for iterative computation substantially slow down its performance on moderately sized datasets, especially for machine learning algorithms. However, Spark is popular as it is efficient at iterative computations and thus well-suited for the development of large-scale machine learning applications (Meng et al., 2015).

Spark is integrated into the existing Big Data Hadoop cluster as a computing engine provided by Oracle cloud. It is also presented solely by Databricks Spark cloud.

In this paper, we developed time-series models to predict the confirmed cases and fatalities using the traditional AROMA & ETS models, and Spark Big Data machine learning, which is linearly scalable when the data set grows.

The paper has the sections of Data Description and the hardware specification for the experiments. Then, the Methodology section illustrates our approach for forecasting. It also presents the experimental results of the time-series algorithms: ARIMA, ETS, Prophet. Then, we offer models to prove whether the Weather affects the number of cases using Random Forrest and Logistic Regression. Finally, it has a Conclusion.

## Data Description

Data is in CSV format and updated daily. It is sourced from this upstream repository maintained by Johns Hopkins University Center for Systems Science and Engineering (CSSE) who have been doing a public service from an early point by collating data from around the world ("COVID-19 dataset", n.d.).

The dataset size is not of over gigabytes but still provides results with efficient standing. As it grows every day in the world, web sites to provide the data set has the issue of Big Data in terms of data volume. Therefore, lately, the data size becomes too big for public web sites such as CSSE to provide as it evolves around 1 GB annually. Our dataset is 300 MB of CSV file format, which is a dataset of the 100 days from Jan 22, 2020 to May 4, 2020. It has 13 columns with a total 8 files having 540 rows in each.

We have cleaned and normalized that data, for example, tidying dates and consolidating several files into normalized time series. We have also added some metadata such as column descriptions and data packaged it.

The collected dataset refers to the cumulative confirmed cases and deaths of COVID-19 that occurred in Los Angeles and New York of the United States from Jan 22 2020 to May 4, 2020. This dataset includes time-series data tracking the number of people affected by COVID-19 worldwide, including:

• Confirmed tested cases of Coronavirus infection.
• The number of people who have reportedly died while sick with Coronavirus.
• The number of people who have reportedly recovered from it.
In order to show if the Weather affects the confirmed cases, we collect the Weather data set from AccuWeather ("AccuWeather", n.d.) and join them with the COVID-19 data set in Spark.

## Methodology

Since related papers are not available on this COVID 19 topic, we have researched existing works that have predictions on other viral diseases like Ebola and the Zika virus. Besides, we have implemented time series forecasting algorithms and models in Azure ML, Databricks, and Oracle Cloud.

Azure Machine Learning Studio enables to build machine learning models quickly with the graphical user interfaces. It has a drag-and-drop interface that doesn't require any coding, and you can add code if you want to. It supports a wide variety of algorithms, including different types of regression. We have written R codes to build models with the algorithms, ARIMA and ETS.

Databricks provides a cloud platform on top of Apache Spark as a unified analytics platform. It is orchestrated with open-source Spark Core with an underlying general execution engine which supports a wide variety of application, Java, Scala and Python API for the ease of development. We implement a time series forecasting model using Facebook Prophet API in Databricks.

Oracle Big Data Computing Edition is a cloud computing service to provide Hadoop and Spark platform. We implement a time series forecasting model to test it. We built classification models for the relationship with the Weather and the number of confirmed cases in Oracle Cloud and Databricks.

## Time Series Forecasting

Time series data is a useful source for information and strategy used in various businesses. Time-series forecasting is for Time Series data (years, days, hours…etc.) for predicting future values using time-series modeling. We implement ARIMA and ETS models as legacy time-series methods. For Big Data time-series predictive analysis, we have used python Facebook prophet in Databricks to predict the number of cases and deaths in Los Angeles and New York. Databricks provides Databricks Runtime for Machine Learning as a ready- to-go environment for machine learning and data science.

## Hardware Specifications

To forecast the confirmed and fatality cases, we have used Microsoft Azure Machine Learning Studio and Databricks community edition to implement Spark ML, which are legacy and Big Data systems, respectively. We have also used the Hadoop spark cluster on the Oracle Cloud platform as another Big Data systems. It is not easy to store and analyze the data set using the legacy systems, which grows daily. Therefore, we developed time series models to predict the confirmed cases and fatalities using Spark Big Data, which is linearly scalable. Thus, even though the data set grows, our solution works without any data volume and performance issues.

The hardware specification is given below:

|  | **Azure** | **Databricks** | **Oracle Cloud** |
|---|---|---|---|
| Memory | < 15 GB | 15.3 GB | 242 GB |
| CPU cores | Unknown | 2 | 32 |
| Number of Nodes | 1 | 1 | 3 |
| Storage | 10 GB | 15GB | 1 TB |
| Programming Language | R | PySpark | PySpark |

**Table 1. Hardware specifications**

## Legacy Forecasting using Azure ML

We have sampled the data set from April 24 2020, to May 4, 2020, which are of about 30 MB, 10 days data set for the legacy systems. The following depicts the use of the traditional time series forecasting in Azure Machine Learning Studio.

**ARIMA model**

ARIMA (Auto Regressive Integrated Moving Average) and ETS (Error Trend and Seasonality, or exponential smoothing) are two of the most commonly used time series forecasting methods with a series of past values. Using the algorithms, you can predict the number of cases and deaths as a time-series forecast.

ARIMA, short for 'Auto-Regressive Integrated Moving Average' is a class of models that explains a given time series based on its pastures, that is, its lags and the lagged forecast errors. ARIMA model is stationary and does not have exponential smoothing counterparts. You can use the model if the past data explains the present data well, which means autocorrelation in the data.

ETS is not stationary and uses exponential smoothing. You can use ETS model if there is a trend and seasonality in the data.

We have implemented an optimal ARIMA model and extend it to Seasonal & Nonseasonal ARIMA using an R programming language in Figures 1 and 2 as below. The 10 days of data set is listed as a time row of 0 to 100.

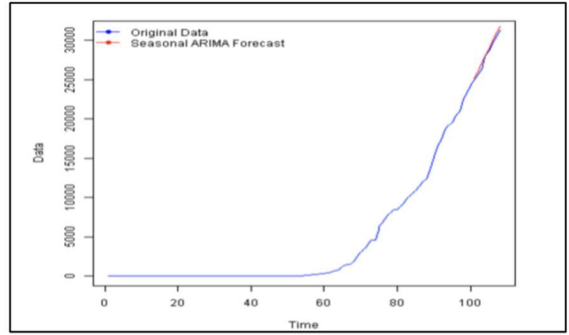

**Figure 1.  Seasonal ARIMA Forecast**

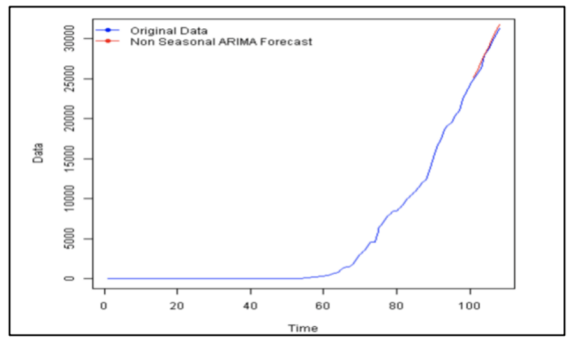

**Figure 2.  Non-Seasonal ARIMA Forecast**

Figures 1 and 2 show the actual and predicted values of the confirmed cases in both seasonal & non-seasonal ARIMA forecasts. The exact RMSE (Root Mean Square Error) is 469 cases, as shown in Table 2, in which accuracy looks good enough because it is 1.5 – 2 % difference from 25,000 to 30,000 cases. The forecast graphs in Figures 1 and 2 look precisely the same. And, it makes sense because we have only 10 days of data set. Therefore, Figure 1 does not show the cycle of seasonal ARIMA.

### ETS model

In Figures 3 and 4, we have implemented Seasonal & Nonseasonal ETS models using an R programming language.

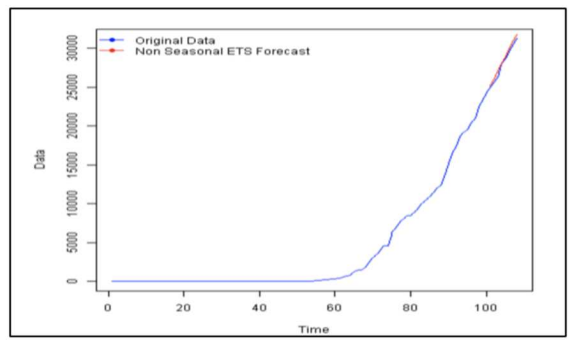

**Figure 3.  Non-Seasonal ETS Forecast**

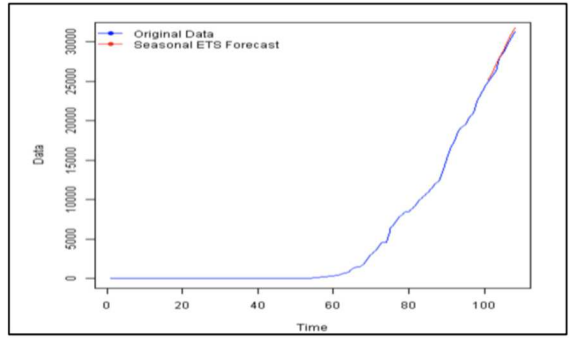

**Figure 4. Seasonal ETS Forecast**

Figures 3 and 4 show the actual and predicted values of the confirmed cases in both seasonal & non-seasonal ETS forecasts. The exact RMSE (Root Mean Square Error) is 472 cases, as shown in Table 2. Therefore, the accuracy looks good because it is 1.5 – 2 % difference from 25,000 to 30,000 cases. The forecast graphs in Figures 3 and 4 look precisely the same, which is similar to the ARIMA forecasts in Figures 1 and 2. It is because we have only ten days of data set. Therefore, Figure 4 does not show the cycle of seasonal ETS, which is similar to the seasonal ARIMA in Figure 2.

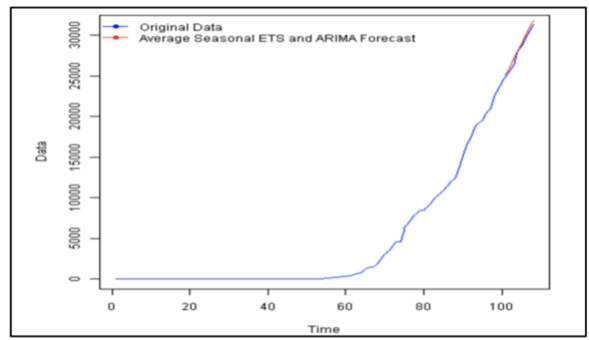

**Figure 5. Average Seasonal ETS and ARIMA Forecast**

Figure 5 shows the average seasonal ETS and ARIMA forecasts, which has the RMSE 471. The actual and predicted cases in average is acceptable.

Table 2 show the accuracies of the models as Root Mean Square Error (RMSE), Mean Error (ME), and Mean Absolute Error (MAE).

| **Azure ML** | **RMSE** | **ME** | **MAE** |
|---|---|---|---|
| Arima Seasonal | 469.1 | -428.2 | 428.2 |
| Arima Non-Seasonal | 469.1 | -428.2 | 428.2 |
| ETS Seasonal | 472.4 | -431.7 | 431.7 |
| ETS Non-Seasonal | 472.4 | -431.7 | 431.7 |
| Average Seasonal ETS and Arima | 470.8 | -429.2 | 429.2 |

**Table 2. Summary- Azure ML Table**

## Big Data Spark Forecast

**Facebook Prophet**

The Facebook prophet is an open-source API for forecasting time series data based on an additive model where non-linear trends are fit with yearly, weekly, and daily seasonality as well as holiday effects. It fits well to the time series with the strong seasonal effects and several seasons of historical data. It is accurate and easy to use, which is built by Facebook's Core Data Science team ("Facebook Prophet", n.d.).

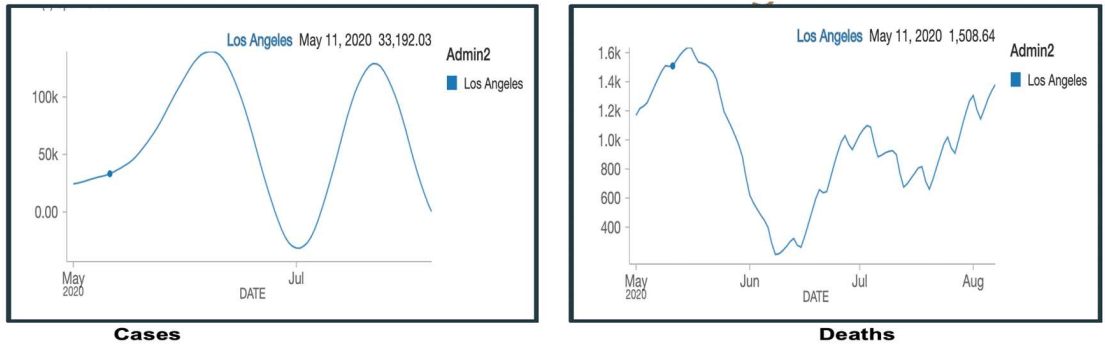

**Figure 6. The Confirmed Cases and Deaths Prediction in LA**

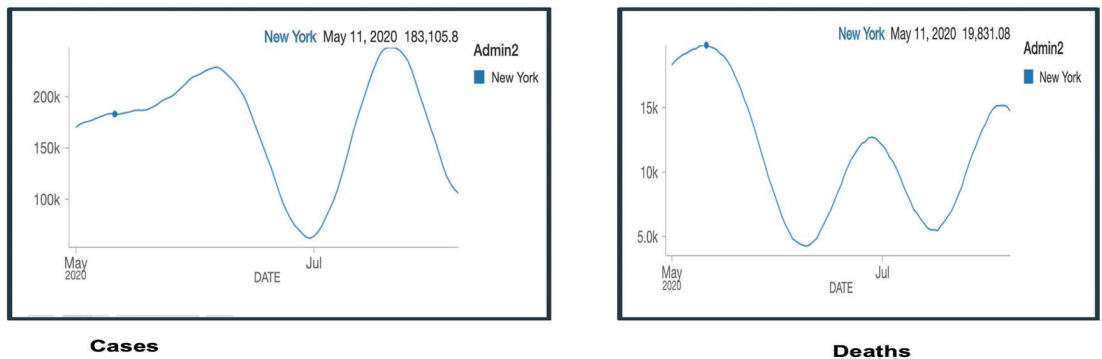

**Figure 7. The Confirmed Cases and Deaths Prediction in NY**

We have used python facebook prophet in Databricks to predict the number of deaths and confirmed cases in Los Angeles and New York, two megacities in the United States where Corona disease is spreading exponentially. The data set is of 300 MB from Jan 22, 2020 to May 4, 2020

Table 3 shows the accuracy of predicting the confirmed cases using Prophet in Los Angeles and New York. RMSE in Los Angeles is 255 and in New York it is 468, which is acceptable while the confirmed cases are around 250 – 300 thousand.

| Facebook Prophet | RMSE | ME | MAE |
|---|---|---|---|
| Los Angeles | 254.9 | 181.6 | 181.6 |
| New York | 467.5 | 335.6 | 335.6 |

**Table 3. Accuracy of Prophet**

## Relation betweenWeather and the Confirmed Cases

**Logistic Regression**

Logistic regression is a statistical algorithm that uses a logistic function to model a binary dependent variable, although many more complex extensions exist.

**Random Forrest Model**

Random Forrest algorithm builds classification models in the form of a tree structure. It breaks down a dataset into smaller and smaller subsets while at the same time, associated trees are incrementally developed. It is an ensemble method to select the best tree model of the number of trees.

| **Spark** | **AUC** |
|---|---|
| Logistic Regression Model | 0.9160 |
| Random Forrest Regression Model | 0.9416 |
| **Oracle BDCE** | **AUC** |
| Logistic Regression Model | 0.9597 |
| Random Forrest Regression Model | 0.9658 |

**Table 3. Correlation of the temperature and the cases**

For determining if *Weather* dominates the number of cases, we implemented Logistic Regression and Random Forrest models using PySpark in Databricks for smaller dataset and Oracle BDCE for the larger dataset, respectively. We defined the field "confirmed cases" as "label" and utilized "Temperature", "Latitude", "Day" as features. When the model generates the prediction, we compare the number of cases with the label; it did not change the results per Weather. Thus, we conclude that Weather did not affect the number of cases and deaths in COVID 19. Table 3 shows the accuracy of the models, and its AUC (Area Under Curve) is 92 – 96 % in Logistic Regression models and 94 – 97 % in Random Forrest models.

## Conclusion

We present time-series forecasting of COVID 19 using the legacy machine learning systems such as the seasonal & non-seasonal ARIMA and ETS models. Besides, we predict the confirmed cases and deaths using Facebook Prophet until August 2020.

We built classification models using Logistic Regression and Random Forrest to find out if the Weather affects the number of confirmed cases. However, it does not show any clue about the relationship.

The experiments are completed with the legacy and the Big Data platforms. We have observed that the data of COVID-19 has grown since December 2019 and the public sites have issues with storing and providing it to the world. Furthermore, the legacy systems cannot process the data as it becomes too large, which causes memory and performance issues. Therefore, we present models suggesting that the Big Data platform using Spark can address the scalability issue.